\documentclass[a4paper]{jpconf}
\usepackage{graphicx}

\usepackage{dsfont}     
\usepackage{subeqnarray}

\newcommand{\be}{\begin{equation}}
\newcommand{\ee}{\end{equation}}
\newcommand{\beq}{\begin{eqnarray}}
\newcommand{\eeq}{\end{eqnarray}}

\newcommand{\I}{{_I}}
\newcommand{\II}{{_{I\!I}}}

\def\boldvec#1{\mbox{\boldmath $#1$\unboldmath}}

\def\lsim{\hbox{ \raise.35ex\rlap{$<$}\lower.6ex\hbox{$\sim$}\ }}
\def\gsim{\hbox{ \raise.35ex\rlap{$>$}\lower.6ex\hbox{$\sim$}\ }}

\newcommand{\bsa}{\begin{subeqnarray}}
\newcommand{\esa}{\end{subeqnarray}}
\newcommand{\bea}{\begin{eqnarray}}
\newcommand{\eea}{\end{eqnarray}}
\newcommand{\ba}{\begin{array}}
\newcommand{\ea}{\end{array}}
\newcommand{\bit}{\begin{itemize}}
\newcommand{\eit}{\end{itemize}}

\def\lab{\label}

\def\lf{\left}

\def\pa{\partial}
\def\ran{\rangle}

\def\ri{\right}

\def\al{\alpha}

\def\ga{\gamma}
\def\Ga{\Gamma}

\def\om{\omega}
\def\Om{\Omega}

\begin{document}

\begin{flushleft}
KCL-PH-TH/2013-3 \\
\end{flushleft}

\title{Noncommutative spectral geometry and the deformed Hopf algebra structure of quantum field theory}

\author{Mairi Sakellariadou} \address{Department of Physics, King's
  College, University of London, Strand WC2R 2LS, London, U.K.}
\ead{mairi.sakellariadou@kcl.ac.uk}
\author{Antonio Stabile}
\address{Dipartimento di Fisica ``E.R. Caianiello'' and Gruppo Collegato I.N.F.N.,
  Universit\`a di Salerno, I-84100 Salerno, Italy}
\ead{anstabile@gmail.com}
\author{Giuseppe Vitiello}
\address{Dipartimento di Fisica ``E.R. Caianiello'' and Gruppo Collegato I.N.F.N., Universit\`a di Salerno, I-84100 Salerno, Italy}
\ead{vitiello@sa.infn.it}

\begin{abstract}
We report the results obtained in the study of Alain Connes noncommutative spectral geometry construction focusing on its essential ingredient of the algebra doubling. We show that such a two-sheeted structure is related with the gauge structure of the theory, its dissipative character and carries in itself the seeds of quantization. From the algebraic point of view, the algebra doubling process has the same structure of the deformed Hops algebra structure which characterizes quantum field theory.
\end{abstract}

\section{Noncommutative spectral geometry and the standard model}
\label{NCSG}
One may assume that near the Planck energy scale, the geometry of space-time ceases
to have the simple continuous form we are familiar with. At high enough energy scales, quantum gravity effects turn on and they alter space-time. One can thus assume that at high energy scales, space-time becomes discrete and the coordinates do not longer commute.
Such an approach could {\sl a priori} be tested by its phenomenological and cosmological consequences.

Combining noncommutative geometry~\cite{ncg-book1,ncg-book2}  with the spectral action principle, led to Noncommutative Spectral Geometry (NCSG),  used by Connes and collaborators~\cite{ccm} in an attempt to provide a purely geometric explanation for the Standard Model (SM)  of electroweak and strong interactions.
In their approach, the SM is considered as a phenomenological model, which dictates the
geometry of space-time so that the Maxwell-Dirac action
functional leads to the SM action.
The model is constructed to hold at high energy scales, namely at unification scale;  to get its low energy consequences which  will then be tested against current data, one uses standard renormalization techniques. Since the model lives at high energy scales, it can be used to investigate early universe
cosmology~\cite{Nelson:2008uy}-\cite{Sakellariadou:2011dk}~\footnote{See the contribution of M.\ Sakellariadou in the same conference.}.
The purpose of this contribution is twofold: firstly, to investigate
the physical meaning of the choice of the {\sl almost} commutative
geometry and its relation to quantization~\cite{PRD}, and secondly to explore
the relation of NCSG with the gauge structure of the theory and with dissipation~\cite{PRD}. We will show that Connes construction is intimately related with the deformed Hopf algebra characterizing quantum field theory (QFT)~\cite{BlasoneJizbaVitiello:2011} and therefore the seeds of quantization are built in such a NCSG construction.

We start by summarizing the main ingredients of NCSG,  composed by
a two-sheeted space, made from the product of a four-dimensional
smooth compact Riemannian manifold ${\cal M}$ with a fixed spin structure, by a discrete
noncommutative space ${\cal F}$
composed by only two points. Thus, geometry is specified by the product of a continuous manifold
for space-time times an internal geometry for the SM. The noncommutative nature of the
discrete space ${\cal F}$ is denoted by a spectral triple $({\cal A, H},
  D)$. The algebra ${\cal A}=C^\infty({\cal M})$ of smooth functions on ${\cal M}$  is an involution of operators on the
finite-dimensional Hilbert space ${\cal H}$ of Euclidean fermions; it acts on ${\cal H}$ by multiplication operators.
The
operator $D$ is the Dirac operator
${\partial\hspace{-5pt}\slash}_{\cal
  M}=\sqrt{-1}\gamma^\mu\nabla_\mu^s$ on the spin Riemannian manifold
${\cal M}$;  $D$ is a self-adjoint unbounded operator in ${\cal H}$. The
space ${\cal H}$ is the Hilbert space $L^2({\cal M},S)$ of square
integrable spinors $S$ on ${\cal M}$. Thus one obtains a model of pure gravity with an action that depends on the spectrum of the Dirac operator, which provides the (familiar) notion of a metric.

The most important ingredient in this NCSG model is the choice of the algebra constructed within the geometry of the two-sheeted space ${\cal M}\times {\cal F}$; it captures all information about space.
The product geometry is specified by:
\beq \lab{1} {\cal A}={\cal A}_1\otimes {\cal A}_2~,~~~~{\cal H}={\cal
  H}_1\otimes {\cal H}_2~,
\eeq
as a consequence of the noncommutative nature of the
discrete space ${\cal F}$ composed by only two points. In other words, Eq.~(\ref{1}) expresses, at the algebraic level and at the level of the space of the states, the two-sheeted nature of the geometry space ${\cal M}\times {\cal F}$.
Assuming ${\cal A}$ is symplectic-unitary, it
can be written as~\cite{Chamseddine:2007ia}
\begin{equation}
\mathcal{A}=M_{a}(\mathds{H})\oplus M_{k}(\mathds{C})~,
\end{equation}
with $k=2a$ and $\mathds{H}$
being the algebra of quaternions, which encodes the noncommutativity of the manifold.   The first possible value for the
even number $k$ is 2, corresponding to a Hilbert space of four
fermions, but this choice is ruled out from the existence of
quarks. The next possible value is $k=4$ leading to the correct number
of $k^2=16$ fermions in each of the three generations. Thus, we will consider the minimal choice that can accommodate the physics of the SM; certainly other choices leading to larger algebras which could accommodate particle beyond the SM sector could be possible.

The second basic ingredient of the NCSG is the spectral
action principle stating that, within the context of the product ${\cal M}\times {\cal F}$, the bare bosonic Euclidean action is given by
the trace of the heat kernel associated with the square of the
noncommutative Dirac operator and is simply
\be {\rm Tr}(f(D/\Lambda))~, \ee
where $f$ is a cut-off function and$\Lambda$ fixes the energy scale. This action can
be seen {\sl \`a la} Wilson as the bare action at the mass scale
$\Lambda$.  The fermionic term can be included in the action
functional by adding $(1/2)\langle J\psi,D\psi\rangle$, where $J$ is
the real structure on the spectral triple and $\psi$ is a spinor in
the Hilbert space ${\cal H}$ of the quarks and leptons.

Since we are considering a four-dimensional Riemannian geometry, the trace ${\rm
  Tr}(f(D/\Lambda))$ can be expressed perturbatively
as~\cite{sdw-coeff}-\cite{nonpert}
\be\label{asymp-exp} {\rm Tr}(f(D/\Lambda))\sim
2\Lambda^4f_4a_0+2\Lambda^2f_2a_2+f_0a_4+\cdots +\Lambda^{-2k}f_{-2k}a_{4+2k}+\cdots~,
\ee
in terms of the
geometrical Seeley-deWitt coefficients $a_n$, known for any
second order elliptic differential operator.
It is important to note that the
smooth even test function $f$, which decays fast at infinity,
appears through its momenta $f_k$:
\beq \nonumber 
f_0 &\equiv& f(0)~,\\  \nonumber
f_k &\equiv&\int_0^\infty f(u) u^{k-1}{\rm
  d}u\ \ ,\ \ \mbox{for}\ \ k>0 ~,\nonumber\\ \mbox
    f_{-2k}&=&(-1)^k\frac{k!}{(2k)!} f^{(2k)}(0)~.  \nonumber
\eeq
Since the Taylor expansion of the cut-off function vanishes at zero, the
asymptotic expansion of Eq.~(\ref{asymp-exp}) reduces to
\be \label{asympt}
{\rm Tr}(f(D/\Lambda))\sim
2\Lambda^4f_4a_0+2\Lambda^2f_2a_2+f_0a_4~.  \ee
Hence, the cut-off function $f$ plays a r\^ole only through its three momenta
$f_0, f_2, f_4$, which are three real parameters, related to the
coupling constants at unification, the gravitational constant, and the
cosmological constant, respectively. More precisely, the first term in Eq.~(\ref{asympt}) which is in
in $\Lambda^4$ gives a
cosmological term, the second one which is in $\Lambda^2$ gives the Einstein-Hilbert
action functional, and the third one which is  $\Lambda$-independent term yields the
Yang-Mills action for the gauge fields corresponding to the internal
degrees of freedom of the metric.

The NCSG offers a purely geometric approach to the SM of particle physics, where the fermions provide the
Hilbert space of a spectral triple for the algebra and the bosons
are obtained through inner fluctuations of the Dirac operator of the
product ${\cal M}\times {\cal F}$ geometry.  The computation of the asymptotic expression for
the spectral action functional results to the full Lagrangian for the
Standard Model minimally coupled to gravity, with neutrino mixing and
Majorana mass terms. Supersymmetric extensions have been also considered.

In this report we closely follow the presentation of Ref.~\cite{PRD} (see also~\cite{Vienna2012}). 
The relation between the algebra doubling and the deformed Hopf algebra structure of QFT are discussed in Section 2; dissipation, the gauge structure and  quantization in Section 3. Section 4 is devoted to conclusions, where we also mention about the dissipative interference phase arising in the noncommutative plane in the presence of algebra doubling.

\section{Noncommutative spectral geometry and quantum field theory}
\label{Sec2}

Our first observation is that the doubling of the algebra, ${\cal A} \, \rightarrow \,{\cal A}_1\otimes {\cal A}_2$ acting on the ``doubled" space ${\cal H}={\cal H}_1\otimes {\cal H}_2$, which expresses the two sheeted nature of the NCSG  (cf. Eq.~(\ref{1})), is a key feature of quantum theories. As observed also by Alain Connes in Ref.~\cite{ncg-book1},
already in the early years of quantum mechanics (QM), in establishing the ``matrix mechanics" Heisenberg has shown that noncommutative algebras governing physical quantities are at the origin of spectroscopic experiments  and are linked to the discretization of the energy of the atomic levels and angular momentum. One can convince himself that this is the case by observing that in the density matrix formalism of QM the coordinate $x(t)$ of a quantum particle is split into two coordinates
$x_+(t)$ (going forward in time) and $x_-(t)$ (going backward in
time). The forward in time motion and the backward in time motion of the density matrix $W(x_{+},x_{-},t)\equiv \langle x_{+}|\rho (t)|x_{-}\rangle = \psi^* (x_{-},t)\psi (x_{+},t)$, where $x_{\pm}=x\pm y/2$, is described indeed by ``two copies" of the Schr\"odinger equation, respectively:
\be i\hbar {\partial \psi (x_{+},t) \over
\partial t}=H_{+}\psi (x_{+},t), \qquad \qquad -i\hbar {\partial \psi^* (x_-,t) \over \partial t}=H_-\psi^*
(x_-,t), \lab{(4a)}
\ee
which can be written as
\be i\hbar {\partial \langle x_+|\rho (t)|x_-\rangle \over \partial t}=
{\hat H}\ \langle x_+|\rho (t)|x_- \rangle, \lab{(5a)} \ee
where $H$ is given in terms of the two Hamiltonian operators $H_{\pm}$ as
\be {\hat H}=H_+ -H_- ~. \lab{(5b)} \ee

The introduction of a doubled set of
coordinates, $(x_{\pm}, p_{\pm})$ (or $(x,p_{x})$ and
$(y,p_{y})$) and the use of the two copies of the Hamiltonian  $H_{\pm }$  operating on the outer product of two Hilbert spaces ${\cal H}_{+} \otimes {\cal H}_{-}$ thus show  that the eigenvalues of ${\hat H}$ are directly
the Bohr transition frequencies $h \nu_{nm}=E_n-E_m$, which are at the basis of the explanation of spectroscopic structure. We have observed elsewhere that the doubling of the algebra is implicit also in the theory of  the Brownian motion of a quantum particle~\cite{Blasone:1998xt} (see also~\cite{PRD} and the references there quoted) and the doubled degrees of freedom are known~\cite{BlasoneJizbaVitiello:2011} to allow quantum noise effect.

It is thus evident the connection with the two sheeted NCSG. Moreover, it has been shown~\cite{PRD} that as a consequence of the algebra doubling Eq.~(\ref{1}) the NCSG construction has an intrinsic gauge structure and is a thermal dissipative field theory. As we will discuss below, this suggests that Connes construction carries in itself the seeds of quantization, namely it is more than just a classical theory construction.

Let us start by  discussing the simple case of the massless fermion and the $U(1)$ local gauge transformation group. We will see how in this case the doubling of the algebra is related to the gauge structure of the theory. Extension to the massive fermion case, the boson case and non-Abelian gauge transformation groups is possible~\cite{Celeghini:1992a,Celeghini:1993a}.
The system Lagrangian  is
\be \hat L = L - {\tilde L} = - \overline{\psi}
\gamma^{\mu}\partial_{\mu}\psi + \overline{\tilde {\psi}}
\gamma^{\mu}\partial_{\mu} \tilde{\psi}. \label{(7)} \ee
The fermion tilde-field $\tilde{\psi}(x)$,
which satisfies the usual fermionic anticommutation relations and anticommutes with the field $\psi(x)$,
is a ``copy"
(with the same spectrum and couplings) of the $\psi$-system. We thus ``double'' the field algebra by introducing such a tilde-field $\tilde{\psi}(x)$.

For simplicity,  no coupling term of the field ${\psi}(x)$ with ${\tilde
{\psi}(x)}$ is assumed in $\hat L$.  In the  quantized theory, let   $a_{\bf k}^{\dag}$ and $\tilde a_{\bf k}^{\dag}$ denote the creation operators associated to the quantum fields $\psi $ and $\tilde {\psi}$,
respectively (all quantum number indices are suppressed except
momentum).
The vacuum
$|0({\theta}) \rangle$ of the theory is:
\be|0(\theta) \rangle  =\prod_k \lf[\cos\theta_k + \sin\theta_k a_{\bf
k}^{\dag} \tilde a_{\bf k}^{\dag}\ri] |0 \rangle  , \label{(g9)} \ee
namely a condensate of couples of $a_{\bf k}^{\dag}$ and $\tilde a_{\bf k}^{\dag}$ modes.
Here $|0 \rangle$ denotes the vacuum $|0, 0 \rangle \equiv |0 \rangle \, \otimes \, |{\tilde 0} \rangle$, with $|0 \rangle$ and $|{\tilde 0} \rangle$ the vacua annihilated by the annihilation operators $a_{\bf k}$ and $\tilde a_{\bf k}$, respectively. On the other hand, $|0(\theta) \rangle  $ is the vacuum with
respect to the fields ${\psi(\theta;x)}$ and $\tilde
{\psi}(\theta;x)$ which are obtained
by means of the Bogoliubov transformation:
\bsa \psi(\theta ; x) &=& B^{-1}(\theta) \psi(x) B(\theta),
\\ [2mm]
\tilde \psi (\theta;x) &=& B^{-1}(\theta) \tilde \psi(x)
B(\theta)~, \label{ex(10)}
\esa
where $B(\theta) \equiv e^{-i{\cal G}}$, with the generator ${\cal G} = -i \sum_{\bf k} \theta_{k} (a_{\bf
k}^{\dag} \tilde a_{\bf k}^{\dag} - a_{\bf
k} \tilde a_{\bf k})$. For simplicity,
${\theta}$ is assumed to be independent of space-time. Extension to space-time dependent Bogoliubov transformations can be done~\cite{Celeghini:1993a}.
$|0({\theta}) \rangle $  is a $SU(2)$ generalized coherent state~\cite{Perelomov:1986tf}.

The Hamiltonian for the $\{\psi(x), \, \tilde{\psi}(x)\}$ system is ${\hat H} = H - {\tilde H}$ (to be compared with Eq.~(\ref{(5b)})), and is given by ${\hat H} = \sum_{\bf k} \hbar\, \om_{\bf k}(a_{\bf k}^{\dag} a_{\bf k} - \tilde a_{\bf k}^{\dag} \tilde a_{\bf k})$. The $\theta$-vacuum $|0(\theta) \rangle$ is  the the zero eigenvalue eigenstate of ${\hat H}$. The relation $ [ a_{\bf k}^{\dag} a_{\bf k} - \tilde
a_{\bf k}^{\dag} \tilde a_{\bf k} ]|0(\theta) \rangle = 0$, for any ${\bf k}$ (and any $\theta$)
characterizes the $\theta$-vacuum structure and it is called the  $\theta$-vacuum condition.

The space of states ${\hat {\cal H}} = {\cal H} \otimes {\tilde {\cal H}} $ is constructed  by repeated applications
of creation operators of ${\psi(\theta;x)}$ and ${\tilde
\psi(\theta;x)}$ on $|0({\theta}) \rangle $ and is called the $\theta$-representation
$\lbrace |0({\theta}) \rangle \rbrace $. $\theta$-representations corresponding to different values of the $\theta$ parameter are unitarily inequivalent representations of the canonical anti-commutation relations in QFT~\cite{BlasoneJizbaVitiello:2011}. The state $|0({\theta}) \rangle $ is  known to be a finite temperature state~\cite{BlasoneJizbaVitiello:2011,Celeghini:1992a,Celeghini:1993a,Celeghini:1998a,Umezawa:1982nv}, namely one finds that $\theta$ is a temperature-dependent parameter. This tells us that the algebra doubling leads to a thermal field theory. The Bogoliubov transformations then induce transition through system phases at different temperatures.

We now consider the subspace
${\cal H}_{\theta c} \, {\subset} \,  \lbrace |0({\theta}) \rangle \rbrace$ made of all the states $|a \rangle_{\theta c} $, including $|0({\theta}) \rangle $,
such that the {\it $\theta$-state condition}
\be [ a_{\bf k}^{\dag} a_{\bf k} - \tilde a_{\bf k}^{\dag} \tilde
a_{\bf k} ] |a \rangle_{\theta c} = 0 ~, \quad \quad {\rm for ~ any} \quad
{\bf k}, \label{ex(11)}\ee
holds in ${\cal H}_{\theta c}$ (note that Eq.~(\ref{ex(11)}) is similar to the Gupta-Bleurer condition defining the physical states in quantum electrodynamics (QED)).  Let $\langle  ... \rangle_{\theta c}$ denote matrix elements in
${\cal H}_{\theta c}$; we have
\be \langle j_{\mu}(x) \rangle_{\theta c}  = \langle \tilde j_{\mu}(x)
\rangle_{\theta c} ,  \label{(12)} \ee
where $j_{\mu}(x) = {\overline \psi}\gamma^{\mu}\psi$ and ${\tilde j}_{\mu}(x) = {\overline {\tilde \psi}}\gamma^{\mu}{\tilde \psi}$. Equalities between matrix elements in
${\cal H}_{\theta c}$, say $\langle A \rangle_{\theta c}  =  \langle B \rangle_{\theta c} $, are denoted by $A\cong B $ and we call them $\theta$-w-equalities ($\theta$-weak
equalities). They are classical equalities since they are equalities among c-numbers. ${\cal H}_{\theta c}$ is invariant under the dynamics
described by $\hat H$ (even in the general case in which
interaction terms are present in $\hat H$ provided that the charge is
conserved).

The key point is that, due to Eq.~(\ref{(12)}), the matrix
elements in ${\cal H}_{\theta c}$ of the Lagrangian Eq.~(\ref{(7)})  are invariant under the simultaneous local gauge
transformations of $\psi$ and $\tilde \psi$ fields given by
\be
\psi(x)\to \exp{[ig\alpha(x)]}\psi(x),
 \qquad \qquad  \tilde \psi(x) \to  \exp{[ig\alpha(x)]}\,  \tilde
\psi(x) ~, \label{(13)} \ee
i.e.,
\be \langle \hat L \rangle_{\theta c}   \to   \langle  \hat L^\prime
\rangle_{\theta c} = \langle \hat L \rangle_{\theta c}~,~~{\rm
in}~~{\cal H}_{\theta c}, \label{ex(14)} \ee
under the gauge transformations (\ref{(13)}).
The tilde term $\overline{\tilde{\psi}}\gamma^\mu\partial_{\mu}\tilde{\psi}$
thus plays a crucial r\^ole
in the $\theta$-w-gauge invariance of $\hat L$ under
Eq.~(\ref{(13)}). Indeed it transforms in such a way to compensate the
local gauge transformation of the $\psi$ kinematical term, i.e.,
\be \overline {\tilde \psi} (x) \gamma^{\mu}\partial_{\mu} \tilde \psi
(x) \to \overline {\tilde \psi} (x) \gamma^{\mu} \partial_{\mu} \tilde
\psi (x) + g \partial^{\mu}\alpha(x) \tilde{j}_{\mu}(x). \label{(15)}
\ee
This suggests to us to introduce the vector field $A_{\mu}^\prime$ by
\be gj^{\bar \mu} (x) A_{\bar \mu}^\prime(x)\cong \overline {\tilde \psi}
(x) \gamma^{\bar \mu}
\partial_{\bar \mu} \tilde \psi (x)~,~~~
\bar \mu=0,1,2,3. \label{(16)} \ee
Here and in the following, the bar over ${\mu}$ means no summation on repeated indices. Thus, the vector field $A_{\mu}^\prime$ transforms
as
\be A_{\mu}^\prime(x) \to A_{\mu}^\prime(x) +
\partial_{\mu}\alpha(x),
 \label{ex(17)} \ee
when the transformations of  Eq.~(\ref{(13)})  are implemented.  In ${\cal H}_{\theta c}$, $A_{\mu}^\prime$ can be then identified with the
conventional U(1) gauge vector field and can be  introduced it in the original Lagrangian through the usual coupling term $ig{\overline \psi}\gamma^{\mu}\psi
A_{\mu}^\prime$.

The position (\ref{(16)}) does not change the $\theta$-vacuum
structure. Therefore, provided that one restricts himself/herself  to matrix
elements in ${\cal H}_{\theta c}$,  matrix
elements of physical observables, which are solely functions of
the ${\psi}(x)$ field, are not
changed by the position (\ref{(16)}).
Our identification of $A_{\mu}^\prime$ with the
U(1) gauge vector field is also justified by the fact that observables turn out to be
invariant under gauge transformations and the conservation laws derivable from $\hat L$, namely in the simple case of Eq.~(\ref{(7)}) the current conservation laws, $\partial^{\mu}j_{\mu}(x) = 0$ and $\partial^{\mu}\tilde j_{\mu}(x) = 0$, are also preserved as $\theta$-w-equalities when Eq.~(\ref{(16)}) is adopted. Indeed, one obtains~\cite{Celeghini:1992a,Celeghini:1993a} $\partial^\mu j_\mu (x) \cong 0$ and $\partial^{\mu} \tilde j_{\mu} (x) \cong 0$.
One may also show that
\be \label{(29)} \partial^{\nu}F_{\mu\nu}^\prime(x) \cong -gj_{\mu}(x), \quad \quad
\partial^{\nu}F_{\mu\nu}^\prime(x) \cong -g \tilde j_{\mu}(x),
 \ee
in ${\cal H}_{\theta c}$. In the the  Lorentz gauge from Eq.~(\ref{(29)}) we also obtain the $\theta$-w-relations $\partial^{\mu}A_{\mu}^\prime(x) \cong 0$ and $\pa^{2} A_{\mu}^\prime(x) \cong gj_{\mu}(x)$.

In conclusion,  the ``doubled algebra" Lagrangian (\ref{(7)}) for the field $\psi$ and its ``double" $\tilde \psi$ can be substituted in
${\cal H}_{\theta c}$ by:
\be \hat L_{\rm g} \cong - 1/4{F^{\prime \mu\nu}} F_{\mu\nu}^\prime - \overline
\psi \gamma^{\mu}\partial_{\mu}\psi + ig{\overline \psi}\gamma^{\mu}\psi
A_{\mu}^\prime ~, ~~{\rm
in}~{\cal H}_{\theta c}~, \label{(31)} \ee
where, remarkably, the
tilde-kinematical term $\overline{\tilde {\psi}}
\gamma^{\mu}\partial_{\mu} \tilde{\psi}$ is replaced, in a $\theta$-w-sense, by the coupling term $ig{\overline \psi}\gamma^{\mu}\psi
A_{\mu}^\prime$ between the gauge field $A_{\mu}^\prime$ and the matter field current ${\overline \psi}\gamma^{\mu}\psi$.

Finally, in the case an interaction term is present in the
Lagrangian (\ref{(7)}), $\hat L_{\rm tot}={\hat L}+{\hat
L}_{I},~~~{\hat L}_{I}=L_{\rm I}-{\tilde L}_{\rm I}$, the above conclusions  still hold provided ${\cal H}_{\theta c}$ is an invariant
subspace under the dynamics described by $\hat L_{\rm tot}$.

Our discussion has thus shown that the ``doubling" of the field algebra introduces a gauge structure in the theory: Connes two sheeted geometric construction has intrinsic gauge properties.

The algebraic structure underlying the above discussion is recognized to be the one of the noncommutative $q$-deformed Hopf algebra~\cite{Celeghini:1998a}. We remark that the Hopf coproduct map
${\cal A} \, \rightarrow \, {\cal A} \otimes \mathds{1} + \mathds{1}
\otimes {\cal A} \equiv \, {\cal A}_1 \otimes {\cal A}_2$ is nothing but
the map presented in Eq.~(\ref{1}) which
duplicates the algebra. On the other hand, it can be shown~\cite{Celeghini:1998a} that the Bogoliubov transformation of ``angle"
$\theta$ relating the fields $\psi (\theta; x)$ and ${\tilde \psi}
(\theta; x)$ to $\psi (x)$ and ${\tilde \psi} (x)$, Eqs.~(\ref{ex(10)}), are obtained by convenient combinations of the $q$-{\it deformed} Hopf coproduct $\Delta a^{\dag}_q=a^{\dag}_q\otimes q^{1/2} +
q^{-1/2}\otimes a^{\dag}_q$, with $q \equiv q (\theta)$ the
deformation parameters and $a^{\dag}_q$ the creation operators in
the $q$-deformed Hopf algebra~\cite{Celeghini:1998a}. These deformed
coproduct maps are noncommutative.  All of this signals a deep physical meaning of noncommutativity in the Connes construction since the deformation parameter is related to the condensate content of
$|0 (\theta) \rangle$ under the constrain imposed by the $\theta$-state condition Eq.~(\ref{ex(11)}). Actually, such a state condition is a characterizing condition for the system physical states.
The crucial point is that a characteristic feature of quantum
field theory~\cite{BlasoneJizbaVitiello:2011,Celeghini:1998a} is that the deformation parameter {\it labels} the
$\theta$-representations $\{|0 (\theta) \rangle\}$ and, as already mentioned, for $\theta
\neq \theta'$, $\{|0 (\theta) \rangle\}$ and $\{|0 (\theta')
\rangle\}$ are unitarily inequivalent representations of the canonical
(anti-)commutation rules~\cite{BlasoneJizbaVitiello:2011,Umezawa:1982nv}. In turn, the physical meaning of this is that an order parameter exists, which assumes different
$\theta$-dependent values in each of the representations. Thus, the $q$-deformed Hopf algebra structure of QFT induces the {\it foliation}
of the whole Hilbert space into physically inequivalent subspaces. From our discussion we conclude that this is also the scenario which NCSG presents to us.

One more remark in this connection is that in the NCSG construction the
derivative in the discrete direction is a finite difference quotient~\cite{ncg-book1,ncg-book2,PRD} and it is then suggestive that the $q$-derivative is also a finite difference derivative. This point deserves further formal analysis which is in our plans to do.

We thus conclude that Connes NCSG construction is built on the same noncommutative deformed Hopf algebra structure of QFT.
In the next Section we show that it is also related with dissipation and carries in it the seeds of quantization.

\section{Dissipation, gauge field and quantization}

In the second equation in (\ref{(29)}) the current $\tilde j_{\mu}$ act as the source of the  variations of the gauge field tensor $F_{\mu \nu}^\prime$. We express this by saying that the tilde field plays the r\^ole of a ``reservoir". Such a reservoir interpretation may be extended also to the gauge field
$A_{\mu}^\prime$, which is known to act, indeed, in a way to ``compensate" the changes in the matter field configurations due to the local gauge freedom.

When we consider variations in the $\theta$ parameter (namely in the $q$-deformation parameter), induced by the Bogoliubov transformation generator, we have (time-)evolution over the manifold of the $\theta$-labeled (i.e. $q$-labeled) spaces and we have dissipative fluxes between the doubled sets of fields, or, in other words, according to the above picture, between the system and the reservoir. We talk of dissipation and open systems when considering the Connes conctruction and the Standard Model in the same sense in a system of electromagnetically interacting matter field, neither the energy-momentum tensor of the matter field, nor that of the gauge field, are conserved.  However, one verifies in a standard fashion~\cite{Landau} that $\partial_{\mu} T^{\mu
  \nu}_{\rm matter} = e F^{\mu \nu} j_{\mu} = - \partial_{\mu} T^{\mu
  \nu}_{\rm gauge\ field}$, so that what it is
conserved is the {\sl total} $T^{\mu \nu}_{\rm
  total} = T^{\mu \nu}_{\rm matter} + T^{\mu \nu}_{\rm gauge
  \ field}$, namely the energy-momentum tensor of the {\it closed}
system \{matter field, electromagnetic field\}. As remarked in Ref.~\cite{PRD},  each element of the couple is {\it open} (dissipating) on the other one, although the {\it closeness} of the total system is ensured. Thus the closeness of the SM is not spoiled in our discussion.

In order to further clarify how the gauge structure is related to the algebra doubling and to dissipation we consider the prototype of a dissipative system, namely the classical one-dimensional damped harmonic oscillator
\be
m \ddot x + \gamma \dot x + k x  = 0~, \label{2.1a}
\ee
and its  time-reversed $(\gamma \rightarrow - \gamma)$ (doubled) image
\be
m \ddot y - \gamma \dot y + k y  = 0 ~,\label{2.1b}
\ee
with time independent $m$, $\gamma$ and $k$, needed~\cite{Celeghini:1992yv}
in order to set up the canonical formalism for open systems. The system of Eq.~(\ref{2.1a}) and Eq.~(\ref{2.1b})
is then a closed system described by the Lagrangian density
\be L (\dot{x},\dot{y},x,y)= m\dot{x}\dot{y}+ {\ga \over
  2}(x\dot{y}-y\dot{x})+k x\,y~. \lab{(26)} \ee
It is convenient to use the coordinates ${{x_1}(t)}$ and ${{x_2}(t)}$
defined by
\be x_{1}(t) = \frac{x(t) + y(t)}{\sqrt{2}}~, \qquad x_{2}(t) =
\frac{x(t) - y(t)}{\sqrt{2}}~. \ee
The motion equations are rewritten then as
\bsa
m \ddot x_1 + \gamma \dot x_2 + k x_1  &=& 0~, \label{2.16}
\\
[1mm]
m \ddot x_2 + \gamma \dot x_1 + k x_2   &=& 0~.
\esa
The canonical momenta are: $\, p_{1} = m {\dot x}_{1} + (1/2) \ga {x_2}$ ; $p_{2} = - m
{\dot x}_{2} - (1/2) \ga {x_1} \,$; the Hamiltonian is
\bea {\hat H} &=& H_1 - H_2  = {1 \over 2m} (p_1 - {\gamma\over
  2}x_2)^2 + {k\over 2} x_1^2  -{1
  \over 2m} (p_2 + {\gamma\over 2}x_1)^2 - {k\over 2}
x_2^2~. \label{2.17} \eea
We then recognize~\cite{Tsue:1993nz,Blasone:1996yh,Celeghini:1992a,Celeghini:1993a} that
\be A_i = {B\over 2} \epsilon_{ij} x_j~, ~~~(i,j = 1,2)~,  \qquad \qquad   {\epsilon}_{ii} = 0~,~~ {\epsilon}_{12}
  = - {\epsilon}_{21} = 1~, \label{2.21}
\ee
with $B \equiv {c\, \gamma/e}$,  acts as the vector potential
and obtain that the system of oscillators Eq.~(\ref{2.1a}) and Eq.~(\ref{2.1b}) is equivalent to the system of two
particles with opposite charges $e_1 = - e_2 = e$ in the (oscillator)
potential $\Phi \equiv (k/2/e)({x_1}^2 - {x_2}^2) \equiv {\Phi}_1
- {\Phi}_2$ with $ {\Phi}_i \equiv (k/2/e){x_i}^{2}$ and in the
constant magnetic field $\boldvec{B}$ defined as $\boldvec{B}=
\boldvec{\nabla} \times \boldvec{A} = - B \boldvec{{\hat 3}}$. The Hamiltonian is indeed
\bea {\hat H} = {H_1} - {H_2}  = {1 \over 2m} (p_1 - {e_1 \over{c}}{A_1})^2
+ {e_1}{\Phi}_1  - {1 \over 2m} (p_2 + {e_2
  \over{c}} A_2)^2 + {e_2}{\Phi}_2~.\label{2.22} \eea
and the Lagrangian of the system can be written
in the familiar form
\bea L &=& {1 \over 2m} (m{\dot x_1} + {e_1
  \over{c}} A_1)^2 - {1 \over 2m} (m{\dot x_2} + {e_2 \over{c}} A_2)^2
 - {e^2\over 2mc^2}({A_1}^2 + {A_2}^2) -
e\Phi \label{2.24i} \nonumber\\ &=& {m \over 2} ({\dot x_1}^2 -
{\dot x_2}^2) +{e\over{c}}( {\dot x}_1 A_1 + {\dot x}_2 A_2) -
e{\Phi}~. \label{2.24} \eea
Note the ``minus" sign of the Lorentzian-like (pseudoeuclidean) metric in
Eq.~(\ref{2.24}) (cf. also Eqs.~(\ref{(5b)}), (\ref{(7)}) and(\ref{2.22})), not imposed by hand, but derived through the doubling of the degrees of freedom and crucial in our
description (and in the NCSG construction).

The doubled coordinate $x_2$ thus acts as the gauge
field component $A_1$ to which the $x_1$ coordinate is coupled, and
{\sl vice versa}. The energy dissipated by one of the two systems is
gained by the other one and viceversa, in analogy to what happens in
standard electrodynamics as observed above.
The picture is recovered of the gauge field as the bath or
reservoir in which the system is embedded~\cite{Celeghini:1992a,Celeghini:1993a}.

Our toy system of harmonic oscillators Eq.~(\ref{2.1a}) and Eq.~(\ref{2.1b}) offers also an useful playground to show how dissipation is implicitly related to quantization.  't~Hooft  indeed has
conjectured that classical, deterministic systems with loss of information might lead to a quantum evolution~\cite{'tHooft:1999gk,erice,'tHooft:2006sy} provided some specific energy conditions are met and
some constraints are imposed. Let us verify how such a conjecture is confirmed for our system described by Eq.~(\ref{2.1a}) and Eq.~(\ref{2.1b}). We will thus show that quantization is achieved as a consequence of dissipation. We rewrite the Hamiltonian Eq.~(\ref{2.17}) as ${\hat H} = H_{\rm \I} - H_{\rm \II}$, with
\bea
&&H_{\rm \I} = \frac{1}{2 \Om {\cal C}} (2 \Om {\cal C}
- \Ga J_2)^2 ~~,~~
H_{\rm \II} = \frac{\Ga^2}{2 \Om {\cal C}} J_2^2\,   \label{split}
\eea
where {\cal C} is the Casimir operator  and $J_2$ is the (second) $SU(1,1)$ generator~\cite{Blasone:1996yh}:
\bea\lab{ca} {\cal C} = \frac{1}{4 \Om m}\lf[ \lf(p_1^2  - p_2^2\ri)+
m^2\Om^2 \lf(x_1^2 -  x_2^2\ri)\ri]~, \qquad
J_2 = \frac{m}{2}\lf[\lf( {\dot x}_1 x_2 - {\dot x}_2
x_1 \ri) - {\Ga} r^2 \ri]
~.
\eea
${\cal C}$ is taken to be positive and
$$\Ga = {\ga\over 2 m}~,~ \Om = \sqrt{\frac{1}{m}
(k-\frac{\ga^2}{4m})}~,~ \mbox{with}~~ k >\frac{\ga^2}{4m}~.$$
This  ${\hat H}$ belongs to the class of Hamiltonians
considered by 't~Hooft and is of the form~\cite{Blasone:2000ew,Blasone:2002hq}
\bea \lab{pqham}
{\hat H} &=& \sum_{i=1}^2p_{i}\, f_{i}(q)\,,
\eea
with $p_1 = {\cal C}$, $p_2 = J_2$, $f_1(q)=2\Om$, $f_2(q)=-2\Ga$. Note that $\{q_{i},p_i\} =1$ and
the other Poisson brackets are vanishing. The $f_{i}(q)$ are nonsingular
functions of  $q_{i}$ and the equations for
the $q$'s, $\dot{q_{i}} = \{q_{i}, H\} = f_{i}(q)$, are
decoupled from the conjugate momenta $p_i$. A complete set of observables, called {\em beables}, then exists, which
Poisson commute at all times.  Thus the system admits a deterministic description even when expressed in
terms of operators acting on some functional space of states
$|\psi\ran$, such as the Hilbert space~\cite{erice}. Note that this  description in terms of operators and Hilbert space, does not
imply {\em per se} quantization of the system. Physical states are defined to be those satisfying the condition  $J_2 |\psi\ran = 0$. This guaranties that ${\hat H}$ is bounded from below and ${\hat H} |\psi\ran= H_{\rm \I} |\psi\ran = 2\Om {\cal C}|\psi\ran$. $H_{\rm \I}$ thus reduces to the
Hamiltonian for the two-dimensional ``isotropic'' (or ``radial'')
harmonic oscillator $\ddot{r} + \Om^2 r =0 $. Indeed, putting $K\equiv m \Om^2$, we obtain
\bea \lab{17}
{\hat H} |\psi\ran= H_{\rm \I} |\psi\ran
= \left( \frac{1}{2m}p_{r}^{2} + \frac{K}{2}r^{2}\right) |\psi \ran \, .
\eea

The physical states are invariant under time-reversal ($|\psi(t)\ran =
|\psi(-t)\ran$) and periodical with period $\tau = 2\pi/\Omega$.
 $ H_{\rm \I} = 2 \Om{\cal C} $ has the spectrum ${\cal
  H}^n_{\rm \I}= \hbar \Om n$, $n = 0, \pm 1, \pm 2, ...$; since ${\cal C}$ has been chosen to be positive, only positive values of
$n$ are considered. Then one obtains
$$\frac{ \langle \psi(\tau)|{\hat H} |\psi(\tau) \rangle }{\hbar} \tau -
\phi = 2\pi n~~,~~ n = 0, 1, 2, \ldots~.$$
Using $\tau = 2
\pi/\Om$ and $\phi = \alpha \pi$, where $\al$ is a real constant, leads to
\bea\lab{spectrum}
{\cal H}_{\rm \I,e\!f\!f}^n \equiv
\langle \psi_{n}(\tau)| {\hat H} |\psi_{n}(\tau) \rangle= \hbar
\Om \lf( n + \frac{\alpha}{2} \ri) ~.
\eea
${\cal H}_{\rm
  \I,e\!f\!f}^n$ gives the effective $n$th energy level of the
system, namely the energy given by ${\cal H}_{\rm \I}^n$
corrected by its interaction with the environment. We conclude that
the dissipation term $J_2$ of the Hamiltonian is responsible for the
zero point ($n = 0$) energy: $E_{0} =(\hbar/2) \Om \alpha$.
In QM the zero point energy {\it is} the ``signature" of
quantization since it is formally
due to the nonzero commutator of the canonically conjugate $q$ and $p$
operators. Our conclusion is thus that the (zero point) ``quantum
contribution" $E_0$ to the spectrum of physical states is due and signals the underlying dissipative dynamics: dissipation manifests
itself as ``quantization".

We remark that the ``full Hamiltonian'' Eq.~(\ref{pqham})
plays  the r{\^o}le of the free energy ${\cal F}$, and $2
\Ga J_{2}$ represents  the heat contribution in ${\hat H}$ (or
$\cal F$).  Indeed, using $S \equiv (2 J_{2}/\hbar)$ and $U \equiv 2 \Om {\cal C}$ and  the defining relation for temperature in thermodynamics: $\pa S/\pa U = 1/T$, with the Boltzmann constant $k_{\rm B} = 1$, from
Eq.~(\ref{pqham}) we obtain $T = \hbar \Ga$, i.e., provided  that $S$ is
identified with the entropy, $\hbar \Ga$ can be regarded as the
temperature. Note that it is not surprising that $2 J_{2}/\hbar$ behaves as
the entropy since it controls the dissipative part of the dynamics (thus the irreversible loss of information).
It is interesting to observe that this thermodynamical picture is
also cofirmed by the results on the canonical quantization of
open systems in quantum field theory~\cite{Celeghini:1992yv}.

On the basis of these results (confirming 't~Hooft's conjecture) we have proved that the NCSG classical construction, due to its essential ingredient of the doubling of the algebra, is intrinsically related with  dissipation and thus carries in itself the seeds of quantization~\cite{PRD}.

\section{Conclusions}

In Ref.~\cite{Sivasubramanian:2003xy} it has been shown that a ``dissipative interference phase'' also appears in the evolution of dissipative systems. Such a feature exhibits one more consequence of the algebra doubling which plays so an important r{\^o}le in the Connes NCSG construction. In this report we have discussed how such a doubling process implies the gauge structure of the theory, its thermal characterization, its built in dissipation features and the consequent possibility to exhibit quantum evolution behaviors. Here we only mention that the doubling of the coordinates also provides a relation between dissipation and noncommutative
geometry in the plane. We refer to Ref.\cite{PRD} for details with reference to Connes construction (see also~\cite{Vienna2012}).
We only recall that in the $ (x_+,x_-)$  plane, $x_{\pm}=x\pm y/2$ (cf.
Section 2), the components  of forward and
backward in time velocity $ v_\pm =\dot{x}_\pm$,  given by
\be v_{\pm }={{\partial  H} \over
\partial p_{\pm }} =\pm \, \frac{1}{m}\lf( p_\pm \mp \frac{\ga}{2}\,
x_\mp \ri) ~, \label{(9)} \ee
do not commute: $[v_+,v_-]=i\hbar \,\ga/ m^2$.
Thus it is impossible to fix these velocities $v_+$ and $v_-$ as being
identical~\cite{Sivasubramanian:2003xy}. Moreover, one may always introduce
a  set of canonical conjugate position coordinates
$ (\xi_+,\xi_-)$  defined by $\xi_\pm = \mp L^2K_\mp$, with $\hbar K_\pm = m v_\pm$, so that $\left[\xi_+,\xi_-  \right] = iL^2$, which
characterizes the noncommutative geometry in the plane $(x_+,x_-)$. $L$ denotes the geometric length scale in the
plane.

One can show~\cite{Sivasubramanian:2003xy} that an Aharonov--Bohm-type phase interference can always be associated with the noncommutative $(X,Y)$ plane where
\begin{equation} \label{1noncomm}
[X,Y]=iL^2~.
\end{equation}
Eq.~(\ref{1noncomm}) implies the
uncertainty relation
$ (\Delta X) (\Delta Y) \ge L^2/2$, due to the zero point fluctuations in the coordinates. Consider now a particle moving in the plane along two paths starting and finishing at the same point, in a
forward and in a backward direction, respectively, forming a closed loop.  The
phase interference $\vartheta$ may be be shown~\cite{Sivasubramanian:2003xy} to be given by $\vartheta = {\cal A}/L^{2}$, where  $ {\cal A} $ is the area enclosed in the closed loop formed by the paths
and Eq.~(\ref{1noncomm}) in the noncommutative plane can be written as
\begin{equation}
[X,P_X]=i\hbar \ \ \ {\rm where} \ \ \ P_X=\left(\frac{\hbar
Y}{L^2}\right)~. \label{phase4}
\end{equation}
The quantum phase
interference between two alternative paths in the plane is thus
determined by the length scale
$ L  $ and the area $ {\cal A} $. In the  dissipative case, it is
$L^2=\hbar / \ga$, and
 the quantum
dissipative phase interference $\vartheta = {\cal A}/L^2 =
{\cal A} \ga/\hbar$ is associated with the two paths in the noncommutative plane, provided $x_+ \neq x_-$. When doubling is absent, i.e., $x_+ = x_-$, the quantum effect disappear. It is indeed possible to show~\cite{Blasone:1998xt} that in such a classical limit case the doubled degree of freedom is associated with ``unlikely processes''  and it may thus be dropped  in
higher order terms in the perturbative expansion, as indeed it happens in Connes construction. At the Grand Unified Theories scale, when inflation took
place, the effect of gauge fields is fairly shielded. However, since these higher order
terms are the ones responsible for quantum
corrections, the second sheet cannot be neglected if one does not want to preclude quantization effects, as it happens once the universe entered the radiation dominated
era.

\end{document}